\documentclass{article}

\usepackage{arxiv}

\usepackage[utf8]{inputenc} 
\usepackage[T1]{fontenc}    
\usepackage{url}            
\usepackage{booktabs}       
\usepackage{amsfonts}       
\usepackage{nicefrac}       
\usepackage{microtype}      
\usepackage{lipsum}		
\usepackage{graphicx}
\usepackage[numbers]{natbib}
\usepackage{doi}
\usepackage[utf8]{inputenc}
\usepackage{amsmath}
\usepackage{multirow}
\usepackage{array}
\usepackage{tabularx}
\newcolumntype{P}[1]{>{\centering\arraybackslash}p{#1}}
\usepackage{booktabs}
\usepackage{rotating}
\usepackage{gensymb}
\usepackage{float}
\usepackage[export]{adjustbox}
\usepackage{color,soul}

\newcommand{\mnsd}[2]{#1\,$\pm$\,#2}

\title{An Assessment of the Eye Tracking Signal Quality Captured in the HoloLens 2}

\author{ \href{https://orcid.org/0000-0002-7656-2662}{\includegraphics[scale=0.06]{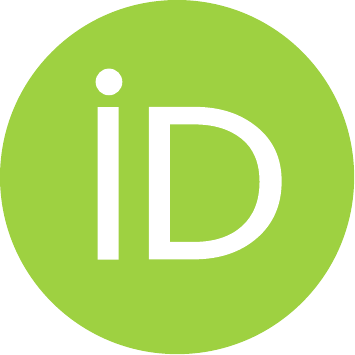}\hspace{1mm}Samantha D.~Aziz}\\
	Department of Computer Science\\
	Texas State University\\
	San Marcos, TX 78666, USA \\
	\texttt{sda69@txstate.edu} \\
	\And
	\href{https://orcid.org/0000-0001-7890-8842}{\includegraphics[scale=0.06]{orcid.pdf}\hspace{1mm}Oleg V.~Komogortsev} \\
	Department of Computer Science\\
	Texas State University\\
	San Marcos, TX 78666, USA \\
	\texttt{ok11@txstate.edu} \\
}

\renewcommand{\shorttitle}{}

\begin{document}
\maketitle

\begin{abstract}
We present an analysis of the eye tracking signal quality of the HoloLens 2’s integrated eye tracker. Signal quality was measured from eye movement data captured during a random saccades task from a new eye movement dataset collected on 30 healthy adults. We characterize the eye tracking signal quality of the device in terms of spatial accuracy, spatial precision, temporal precision, linearity, and crosstalk. Most notably, our evaluation of spatial accuracy reveals that the eye movement data in our dataset appears to be uncalibrated. Recalibrating the data using a subset of our dataset task produces notably better eye tracking signal quality. 
\end{abstract}


\section{Introduction}
The growing ubiquity of dedicated eye trackers into augmented-reality (AR) devices promotes the application of eye tracking to improve interaction between the user and the device. Eye tracking can, for example, prolong the battery life of untethered devices without reducing the perceived quality of the simulated environment through foveated rendering. However, each potential application of eye tracking requires different levels of signal quality to be effective.  Eye movement biometrics \citep{lohr_biometrics}, for example, requires higher quality gaze data than simple gaze-based interaction with the environment \citep{Rajanna2018}. It is therefore imperative to determine whether an eye tracker captures gaze data at a suitable quality for a particular use case.

Data quality metrics such as spatial accuracy and precision are commonly reported by the eye tracking manufacturer. 
However, discrepancies between these values and those achieved in practice have been well-documented in the literature \citep{Rajanna2018, Hornof2002, Holmqvist2017}.
Eye tracking signal quality also depends heavily on the conditions in which the data were collected \citep{Feit2017}. It can be difficult routinely achieve optimal results regardless of experimental condition \citep{Holmqvist2017}. Taken together, these findings underscore the importance of studying eye tracking signal quality under a variety of experimental conditions.

The recently released HoloLens 2 includes a built-in eye tracker, and represents an accessible option for integrating eye tracking into research. In order to benchmark its performance, we evaluate the quality of the eye tracking data captured by the HoloLens 2.
We seek to establish a benchmark for the HoloLens 2's eye tracking signal quality in an experimental setup similar to those commonly found in literature (e.g., \citep{lohr2019evaluating, Griffith2021}), while additionally comparing our results to a previous study of the HoloLens 2's eye tracking signal quality presented by \citep{arett}.
We investigate a number of phenomena---namely, linearity and crosstalk---that exhibited unique properties in an eye movement dataset previously collected in virtual reality~\cite{lohr2019evaluating}. In presenting this analysis, we seek to clarify whether the observations in this virtual reality-based investigation persist across head-mounted virtual/augmented reality eye tracking devices---if they do, these findings would have significant implications for future researchers interested in using eye tracking in virtual and augmented reality.

We introduce a new, publicly available eye movement dataset collected with the HoloLens 2's integrated eye tracker (n=30) and describe its eye tracking signal quality using spatial accuracy, spatial precision, temporal precision, linearity, and crosstalk. From this analysis, we identify discrepancies between the manufacturer-specified signal quality and our own results. We also observe how these results respond to post hoc recalibration. 
The gaze data described in this manuscript is publicly available and can be downloaded at \url{https://doi.org/10.18738/T8/9T99DU}

\section{Methodology}
\subsection{Subjects}
33 subjects (15 female, 18 male, median age: 21, age range: 19-36) took place in this study. Ten subjects who normally wore glasses removed them for this study, and four subjects wore contact lenses while participating. None of the participants wore glasses during data collection.

Three subjects were excluded from this analysis based on excessively noisy data or data loss. Namely, subjects were excluded if more than 20\% of their data consisted of invalid samples. The remaining 30 subjects were used for our signal quality analysis. All subjects wore masks throughout the experiment, as data collection was conducted during the COVID-19 pandemic.

\subsection{Stimulus and Data Collection Procedure}
\label{sec:stimulus}
We employed data from a random saccades task to compute the eye tracking quality measures reported herein. The stimulus shown was a white ring measuring 0.5\degree~displayed at a viewing distance of 1500 mm.~Subjects were instructed to fixate on the center of the stimulus as it appeared in random positions uniformly sampled from a visual field spanning $\pm$15\degree~horizontally and $\pm$10\degree~vertically. The stimulus remained at each position between 1 and 1.5 seconds before jumping to a random location at least 3\degree~away. The stimulus jumped 80 times in each trial.

\begin{figure}
  \begin{center}
    \includegraphics[width=0.6\linewidth]{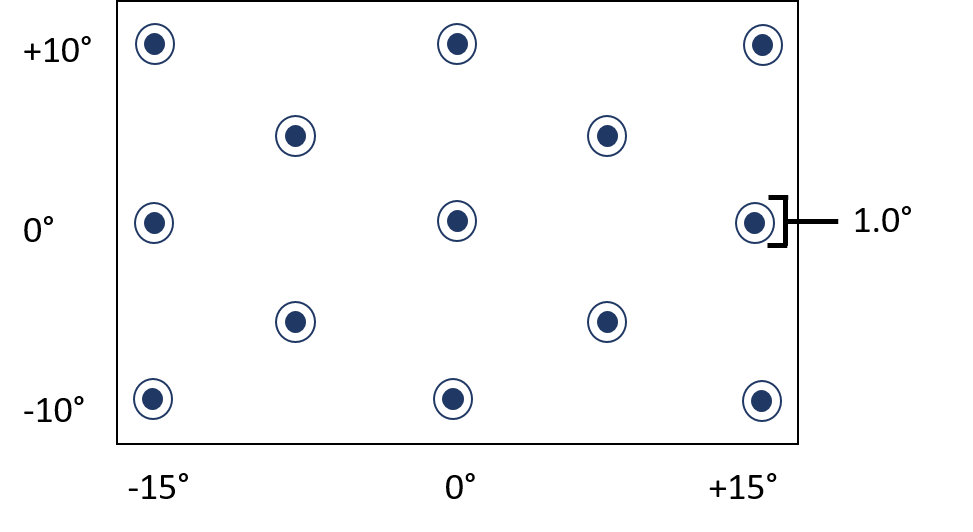}
    \caption{Visualization of our recalibration task.}
    \label{fig:grid}
\end{center}
\end{figure}

To study the effects of manual calibration on the collected data, we also included a recalibration task that took place immediately before the random saccades task. Subjects fixated on a 1\degree~wide stimulus as it appeared in predetermined positions forming a 13-point grid spanning the same field of view as the random saccades task. The positions of these points were the same across all subjects, but were presented in random order. The recalibration task is illustrated in Figure~\ref{fig:grid}.

We created our stimulus using Unity 2019.4.21 and captured gaze data with Microsoft's Mixed Reality Toolkit (MRTK) plugin. We added a custom event to the MRTK to capture and save gaze data reported by the HoloLens 2. Headset position tracking was disabled so that the stimulus remained at a fixed position relative to the center of the headset. To reduce the number of visual distractions from the environment, subjects faced a non-reflective black canvas for the duration of the experiment. Subjects also used a chinrest to minimize head movements during data collection.

\subsection{Eye tracking hardware}
\label{sec:hardware}
The HoloLens 2's integrated eye tracker---referred to herein as the ``HoloLens 2''---features a sampling rate of 30 Hz and a nominal spatial accuracy of 1.5\degree~\citep{MicrosoftEyeTracking}. The HoloLens 2 captures gaze data from the left and right eyes simultaneously. However, the manufacturer does not provide an API to extract raw monocular signals from the device. The two monocular signals are instead combined into a single gaze ray, which is then made available via the MRTK. The methods used to compose this ray from the monocular signal are not publicly available at the time of writing. 

Each gaze sample is represented by a three-dimensional gaze vector $v = (v_x, v_y, v_z)$ vectors. We transformed raw gaze data into degrees of visual angle using simple trigonometry and MATLAB's \texttt{atan2d} function:
\begin{subequations}
\begin{align}
x = \texttt{atan2d}(v_x,  v_z)
\label{eqn:con-x}\\
y = \texttt{atan2d}(v_y,  v_z)
\label{eqn:con-y}
\end{align}
\end{subequations}

\label{sec:calibration}
\subsection{Recalibrating Data}
During data collection, we observed that the HoloLens 2's gaze data appeared to be largely uncalibrated. Gaze positions were systematically offset within subjects, despite calibration taking place immediately before the task began. While the source of this phenomenon is unclear, it is present in all subjects.
Because the HoloLens 2's built-in calibration does not provide feedback, we used the recalibration task described in Section \ref{sec:stimulus} to test the extent to which the captured signal could be calibrated manually.
We selected linear regression as a form of post-hoc data recalibration:
\begin{subequations}
\begin{align}
x' = A_xx + B_xy + C_x
\label{eqn:reg-x}\\
y' = A_yx + B_yy + C_y
\label{eqn:reg-y}
\end{align}
\end{subequations}
where recalibration coefficients \textit{A, B} and \textit{C} were computed using the data obtained from the recalibration task that preceded the random saccades task, as described in Section \ref{sec:stimulus}.
These coefficients were then applied to the data from the random saccades task.

\subsection{Identifying Stable Fixations}
\label{sec:stable}
Eye tracking signal quality is typically measured during a stable fixation on a target. Although there are a number of algorithms available for classifying eye movement (e.g. I-VT), they require input parameters that require fine-tuning, and produce significantly different outputs for non-ideal parameter values \citep{Salvucci2000, Komogortsev2010}. 
Rather than relying on a fixation detection algorithm, used a data-driven approach to select the most stable subset of samples across all subjects.

\begin{figure}
    \begin{center}
        \includegraphics[width=0.6\linewidth]{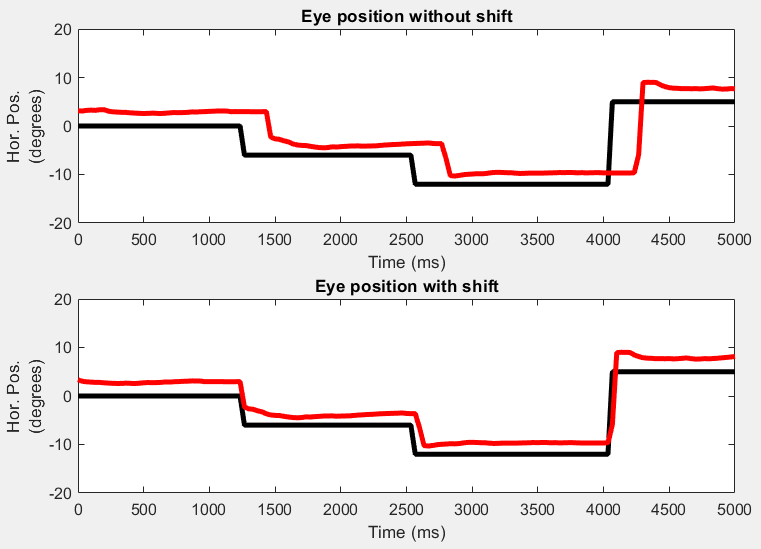}
            \caption{Calculation and removal of saccade latency. The top panel illustrates a typical gaze signal. The bottom panel shows the data after the gaze position signal has been shifted to remove saccade latency.}  
    \label{fig:saccade-latency}
    \end{center}
\end{figure}

First, we minimized the latency between the gaze signal and its corresponding target position using an approach proposed by \cite{lohr2019evaluating}. We estimate the optimal saccade latency for each recording by shifting the gaze signal backward until the mean Euclidean distance between the measured gaze position and the target position was minimized, up to 800 milliseconds. The results of our approach are illustrated in Figure \ref{fig:saccade-latency}.

\begin{figure}
    \begin{center}
        \includegraphics[width=0.6\linewidth]{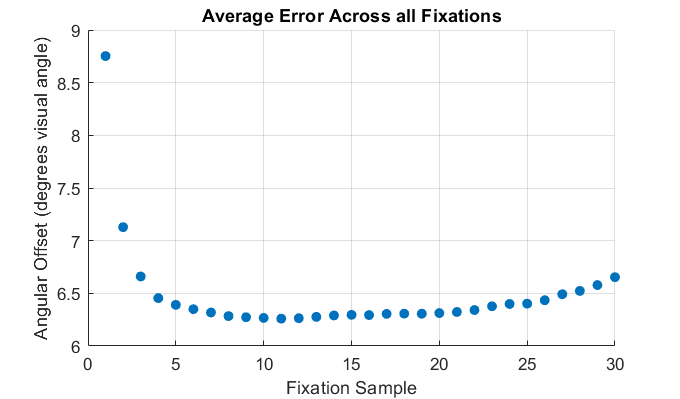}
            \caption{The average sample-by-sample error across all fixations. Samples with the lowest error (8 through 22) were empirically identified as the most stable portion of the signal on average.}
    \label{fig:avg-error}
    \end{center}
\end{figure}

We then identified stable fixation periods by the absence of error due to saccadic movement. 
First, the angular offset between the gaze signal and the target was calculated on a per-sample basis for the first 30 samples (approximately 1000 ms) from the beginning of each target step.
We then calculated the mean angular error of each sample across all fixations and empirically selected the largest contiguous window of samples with the lowest error (Figure \ref{fig:avg-error}), as lower error likely indicates that the eye is stable and fixating on the target \citep{Friedman2021}.
As a result, the first 233 ms of each fixation were discarded to eliminate instability caused by non-fixational movement.
The first 466 ms of the remaining gaze signal was then used for all analysis herein.

\section{Results}
\subsection{Spatial Accuracy}
Spatial accuracy is measured as the distance between the reported gaze position and the target position, reported in degrees of visual angle. 
By treating each stable fixation as a series of \textit{n} gaze samples with a measured gaze position $(x_i^g,y_i^g)$ and a target position $(x_i^t,y_i^t)$, we determine the spatial accuracy of that series with one of the following:
\begin{subequations}
\begin{align}
\theta_H &= \frac{1}{n} \sum_{i=1}^n{|x^g_i - x^t_i|} \label{eqn:acc_h}\\
\theta_V &= \frac{1}{n} \sum_{i=1}^n{|y^g_i - y^t_i|} \label{eqn:acc_v}\\
\theta_C &= \frac{1}{n} \sum_{i=1}^n{\sqrt{(x^g_i - x^t_i)^2 + (y^g_i - y^t_i)^2}} \label{eqn:acc_c}
\end{align}
\end{subequations}
where H, V, and C respectively denote horizontal, vertical, and combined spatial accuracy.
We then take the median spatial accuracy value across fixations within each subject, and present average spatial accuracy of the dataset as the median value across all subjects in the dataset. 
We also include the mean spatial accuracy of our dataset to compare our results with~\cite{arett}'s eye tracking signal quality analysis. 

\begin{figure}
\centering
    \includegraphics[width=0.7\textwidth]{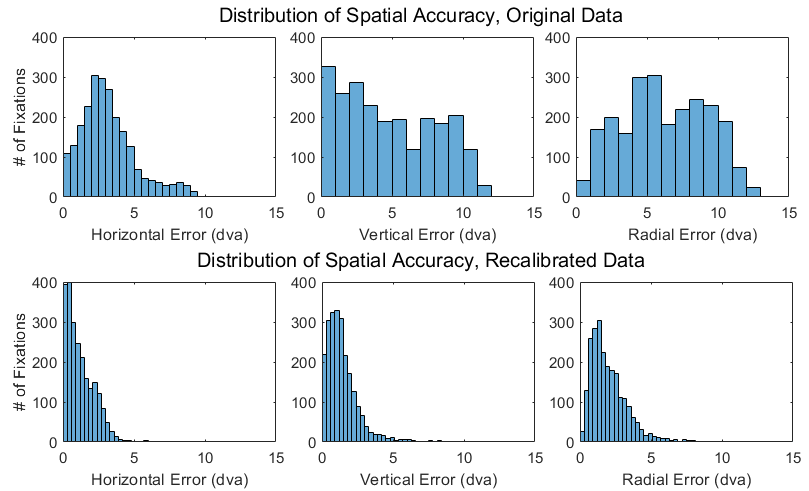}
    \caption{Distribution of spatial accuracy across all measured fixations in the dataset.}
    \label{fig:acc-dist}
\end{figure}

Table \ref{table:accuracy-table} summarizes the spatial accuracy values measured across the dataset. In the original data, accuracy is notably worse in the vertical direction.
The histograms in Figure \ref{fig:acc-dist} show the distribution of spatial accuracy across all fixations in the dataset. 
While the original data has an unusually high frequency of fixations with large error, the recalibrated data closely resembles the right-skewed distribution that is characteristic of spatial accuracy \citep{Holmqvist2017}. 
This corrective effect is seen most prominently in the vertical dimension.

\begin{table}
\centering
\begin{tabular}{@{}ccccP{2cm}P{2cm}@{}}
\toprule
 
\multirow{2}[2]{*}{Calibration} &
\multirow{2}[2]{*}{Dim*} &
\multicolumn{4}{c}{Spatial Accuracy (\degree)} \\
\cmidrule{3-6}

{} & {} & {Median} & {75th Percentile} & {90th Percentile} & {Mean} \\

\midrule

\parbox[t]{3.0cm}{\multirow{3}[3]{*}{Ours (Original)}} &
H &
2.92 &
3.87 &
6.10 &
3.15 \\
{} &
V &
4.16 &
6.86 &
11.09 &
4.94 \\
{} &
C &
6.06 &
7.86 &
12.71 &
6.45 \\
\cmidrule{2-6}

\parbox[t]{3.0cm}{\multirow{3}[3]{*}{Ours (Recalibrated)}} &
H &
0.84 &
1.87 &
2.67 &
1.52 \\
{} &
V &
1.38 &
2.07 &
4.28 &
1.74 \\
{} &
C &
2.06 &
2.90 &
4.94 &
2.66 \\
\cmidrule{2-6}

\parbox[t]{3.0cm}{\multirow{1}[1]{*}{Kapp et al. \citep{arett}}} &
C &
- &
- &
- &
0.77 \\

\bottomrule
\end{tabular}
\caption{Average spatial accuracy across subjects, expressed separately as the 50th, 75th, 90th percentiles, and mean. H, V, and C respectively denote horizontal, vertical, and combined gaze directions.}
\label{table:accuracy-table}
\end{table}

\subsection{Spatial Accuracy Using Vector Comparison}
As a sanity check, we re-calculated the average spatial accuracy for the original data using the raw gaze vectors reported by the HoloLens 2. Given the raw gaze vectors \textit{v} described in Section \ref{sec:hardware} and a target vector \textit{u}, we measured the difference between them using cosine similarity:
\label{eqn:vector-acc}
\begin{align}
\theta = \arccos{\frac{\vec{v} \cdot \vec{u}}{||v|| ||u||}}
\end{align}
The resulting spatial accuracy $\theta$ was nearly identical to the results described in Table \ref{table:accuracy-table}. These results indicate that our data processing and selection efforts did not introduce the extremely high error observed in the original data. 
 
\subsection{Spatial Precision}
\begin{figure}
\centering
    \includegraphics[width=0.4\linewidth]{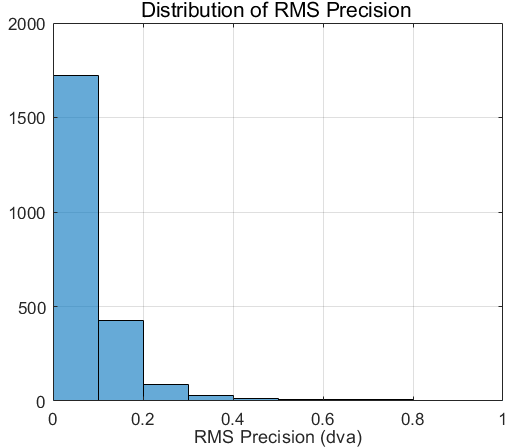}
    \caption{Distribution spatial precision across all measured fixations in the dataset. Spatial precision is similar for the original and recalibrated data, so only the distribution of the original data is shown here.}
    \label{fig:precision-distribution}
\end{figure}

By treating each fixation as a set of \textit{n} gaze samples, we compute spatial precision with the following equation from \cite{Holmqvist}, 

\label{eqn:precision}
\begin{subequations}
\begin{align}
\theta_{RMS} = \sqrt{\frac{1}{n}\sum_{i=1}^n(\theta_i)^2}
\end{align}
\end{subequations}
where $\theta$ is the Euclidean distance between consecutive samples. Although \cite{Holmqvist} recommend using angular distance over Euclidean distance as a measure of intersample distance, \cite{Friedman2021} demonstrate that RMS precision does not differ significantly between the two when calculated on stable fixation periods. Similar to our description of spatial accuracy, we compute the median spatial precision within each subject and describe the average spatial precision of the dataset as the median precision value across subjects. For comparability, we also include the mean precision calculated as the standard deviation of intersample distances described in \citep{arett}.
Table \ref{table:precision-table} summarizes our results. The distribution of precision values across all fixations in the dataset is also shown in Figure \ref{fig:precision-distribution}.

\begin{center}
\begin{table}
\centering
\begin{tabular}{@{}ccccP{2cm}P{2cm}@{}}
\toprule
 
\multirow{2}[2]{*}{Calibration} &
\multirow{2}[2]{*}{Dim*} &
\multicolumn{4}{c}{Spatial Precision (\degree)} \\
\cmidrule{3-6}

{} & {} & {Median} & {75th Percentile} & {90th Percentile} & {Mean$\dagger$} \\

\midrule

\parbox[t]{3.0cm}{\multirow{3}[3]{*}{Ours (Original)}} &
H &
0.071 &
0.086 &
0.011 &
0.15\\
{} &
V &
0.059 &
0.076 &
0.091 &
0.13 \\
{} &
C &
0.06 &
0.09 &
0.11 &
0.14 \\

\cmidrule{2-6}

\parbox[t]{3.0cm}{\multirow{3}[3]{*}{Ours (Recalibrated)}} &
H &
0.065 &
0.082 &
0.102 &
0.14 \\
{} &
V &
0.053 &
0.076 &
0.109 &
0.12 \\
{} &
C &
0.058 &
0.082 &
0.127 &
0.13 \\

\cmidrule{2-6}

\parbox[t]{3.0cm}{\multirow{1}[1]{*}{Kapp et al. \cite{arett}}} &
C &
- &
- &
- &
0.24 \\

\bottomrule
\end{tabular}
\label{table:precision-table}
\caption{Average RMS spatial precision across subjects, expressed separately as the 50th, 75th, and 90th percentiles. H, V, and C respectively denote horizontal, vertical, and combined gaze directions. \textbf{Mean} spatial precision is reported as the standard deviation of intersample distance for comparability. All other precision values are calculated using RMS.}
\end{table}
\end{center}

\subsection{Temporal precision}
We evaluate temporal precision of the device by calculating the variability of inter-sample intervals (ISIs) between consecutive timestamps. Given \textit{n} gaze samples captured by the HoloLens 2 that are each sampled at a timestamp \textit{$t_i$} reported by the device, ISIs are calculated by taking the difference between the timestamps of consecutive samples.
After processing the data captured by the device, we were left with 89,721 gaze samples across all subjects. The mean difference between timestamps is 34.8 ms (SD 21.9 ms), which corresponds to approximately 1 sample. We also investigated the proportion of samples that were dropped by the HoloLens 2. Dropped samples are identified when ISIs exceed 49.95 ms (50\% more than the ideal ISI of 33.3 ms). 4,088 samples across the entire dataset fit this criteria (4.6\% of the dataset).

\subsection{Linearity}
 Linearity measures the extent to which spatial accuracy changes based on the location of the target. Many eye tracking signal quality investigations \citep{Hornof2002}, including those in virtual reality \citep{lohr2019evaluating}, have revealed that spatial accuracy varies systematically relative to the region of the screen the user is looking, typically expressed as the target's location in the field of view. We apply the approach used by~\cite{lohr2019evaluating} for calculating linearity to our own dataset.
 
 \begin{figure}
 \centering
    \includegraphics[width = 0.75\linewidth]{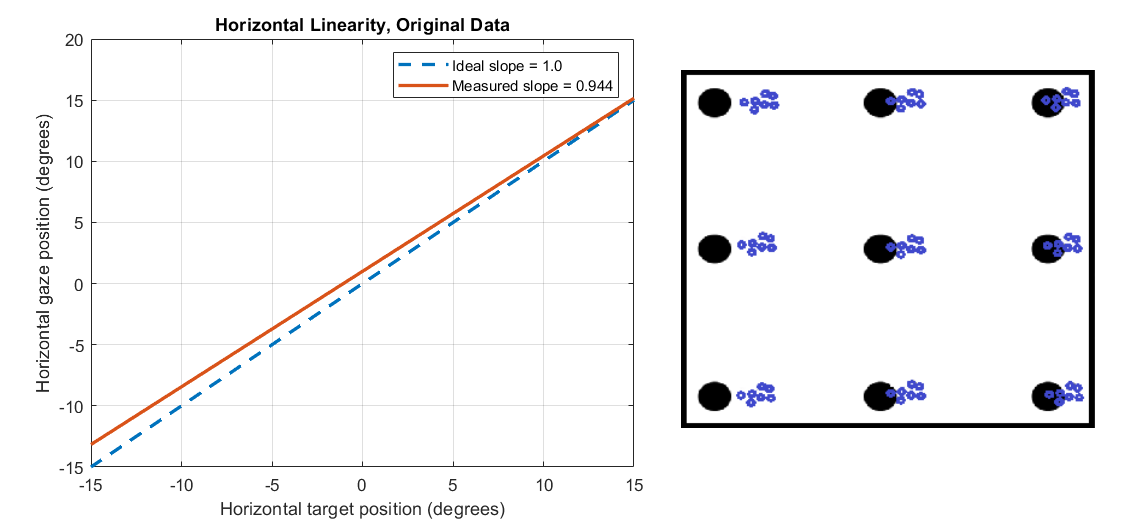}
      \caption{Horizontal linearity results for the original gaze data, with an illustration of the expected accuracy based on the given slope. Gaze samples (blue) would ideally exhibit uniform accuracy across space. Vertical linearity not illustrated.}
    \label{fig:linearity}
\end{figure}
 
 Table \ref{table:lin-table} summarizes the slope of the fitted linearity equation for the horizontal and vertical components of the data across calibration strategies. The ideal linearity slope has a value of 1.0, which indicates that target locations and gaze positions have a straight-line linear association. 
 We identify a linearity slope that is significantly different from ideal when its 95\% confidence interval does not contain the ideal value of 1.0 within its range. 
 While the original data is significantly different from ideal, the recalibrated data features lower error across the field of view and a linearity slope that successfully approximates ideal conditions. To illustrate how linearity is expressed in our own dataset, see Figure~\ref{fig:linearity}.

 \begin{center}
 \begin{table}
 \centering
 \setlength{\tabcolsep}{3pt}
\begin{tabular}{cccc}
\toprule
 
Calibration & Dim &  Linearity Slope & 95\% confidence interval \\
\midrule

\parbox[t]{1.5cm}{\multirow{2}[3]{*}{Original}} &
\multirow{1}{*}{H} &
\mnsd{0.944}{0.050} & [0.925, 0.963] \\

{} &
\multirow{1}{*}{V} &
\mnsd{0.945}{0.043} & [0.929, 0.961] \\

\midrule

\parbox[t]{1.5cm}{\multirow{2}[3]{*}{Recalibrated}} &
\multirow{1}{*}{H} &
\mnsd{0.988}{0.065} & [0.963, 1.012]\\

{} &
\multirow{1}{*}{V} &
\mnsd{0.965}{0.130} & [0.917, 1.014]\\

\bottomrule
\end{tabular}
\label{table:lin-table}
\caption{Linearity slopes by calibration, dimension, and session. Note that linearity is not calculated for combined gaze signals. }
\end{table}
\end{center}

\subsection{Crosstalk}
\begin{figure*}[!htbp]
\centering
\includegraphics[width=\textwidth]{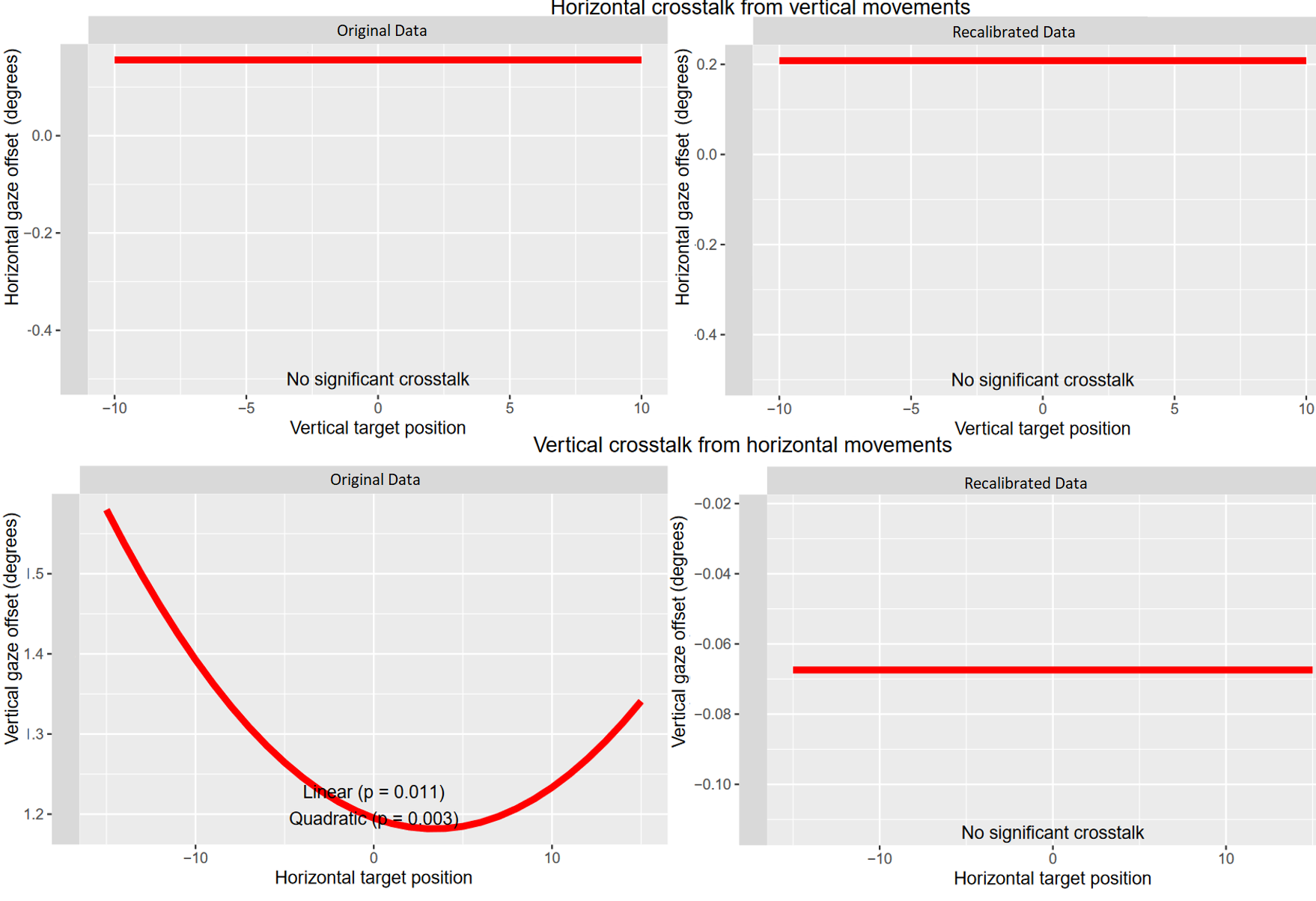}
  \caption{Best-fitting crosstalk models across all subjects in original and recalibrated data measured separately in the horizontal and vertical dimensions. The p-values shown are the significance of each linear or quadratic component.}
    \label{fig:crosstalk}
\end{figure*}

Crosstalk describes the extent to which the rotation of the eye in one direction (e.g. horizontal or vertical) affects movement in the orthogonal direction. Because~\cite{lohr2019evaluating} observed quadratic crosstalk in their virtual reality eye movement dataset, we reproduce their analysis to assess whether this phenomenon is common in virtual/augmented reality eye movement datasets.

Figure \ref{fig:crosstalk} presents the results of crosstalk analysis for our data. Our data does not exhibit significant crosstalk in the horizontal direction in either original or recalibrated data. 
Our evaluation of vertical crosstalk reveals that the best-fitting model varies by calibration method. Vertical crosstalk in the original data has both a linear and quadratic component, which does not persist after recalibration takes place. 

\section{Discussion}
Of all the eye tracking signal quality metrics we investigated, only spatial accuracy is publicly benchmarked by the manufacturer. 
We find that the original data's spatial accuracy results are consistently worse than the typical spatial accuracy reported by the manufacturer (1.5\degree). 
Similarly, \cite{arett} report a mean spatial accuracy of 0.77\degree~in an experimental setup similar to our own---significantly lower (better) than both our findings and the 1.5\degree~accuracy reported by the HoloLens 2 manufacturer. 
Recalibrating our data remarkably improves spatial accuracy.
In fact, the average spatial accuracy achieved by the recalibrated data better approximates the manufacturer's reported spatial accuracy benchmark. 

Although there are no manufacturer-supplied metrics for spatial precision, \cite{arett} provide a benchmark in their own study of HoloLens 2's signal quality, reporting a mean spatial precision of 0.24\degree. 
Our comparable spatial precision results are lower (better) than the prior work at 0.14\degree. The difference in measured spatial precision values may come down to differences in experimental setups---we made efforts to eliminate noise in the data caused by head movement, whereas Kapp et al. did not.

Because crosstalk and linearity are closely related to spatial accuracy, these measures were also improved by recalibration.
Our results for linearity illustrate that spatial accuracy tends to deteriorate at the extremes of the field of vision. This is consistent with the findings of other eye movement signal quality analyses in literature~\cite{Hornof2002, lohr_biometrics}. Recalibration predictably brought linearity values closer to the ideal. The scale of this improvement in linearity may partially depend on the design of the recalibration task, which captures gaze data across the entire field of view. 

Our analysis of crosstalk is based on a novel approach from \citep{lohr2019evaluating}, where quadratic crosstalk was observed in an eye movement dataset captured in virtual reality.
A subset of our data also exhibits partially quadratic crosstalk, but transforms into an intercept-only fit after recalibration. 
It is unclear whether the quadratic vertical crosstalk we initially observed is indeed endemic to head-mounted virtual/augmented reality eye tracking devices, or is simply a consequence of less-than-ideal eye tracking signal quality.
Incidentally, less-than-ideal eye tracking signal quality may be more common in head-mounted devices, as it can be more difficult to consistently achieve a good headset fit on participants.
Future investigations may be able to disentangle the possible sources of crosstalk found in these datasets by grouping individual recordings by spatial accuracy and then analyzing crosstalk separately for each group.

Overall, the recalibrated signal quality results represent a realistic analysis of the HoloLens 2's achievable eye tracking signal quality.
Based on our investigation, it appears that the device's built-in calibration was not applied during our investigation. 
It is also possible that the presence of masks during data collection may be partially responsible for the degradation of eye tracking signal quality in our dataset by introducing spurious corneal reflections or contributing to lens fogging.
However, the relatively lower (higher-quality) spatial precision results our data achieves indicate that this may not be the case.
Our study may have elicited vergence-accommodation conflict in some participants, as we placed our stimulus at a distance of 1.5 m away from the viewer, while the manufacturer recommends a focal plane of 2.0 m in their evaluations of spatial accuracy. 

Further investigation into the eye tracking signal quality of the device is limited by the nature of the gaze data available through the HoloLens 2's eye tracking API.
The device processes gaze data to remove personally identifying information~\citep{MicrosoftEyeTracking}, but it is not known how the signal is further affected (e.g., filtering). 
It is important to note that updates to the manufacturer's eye tracking API since the time of writing may change the characteristics of the gaze signal that is presented to users.

\section{Conclusion}
We evaluated the HoloLens 2's eye tracking signal quality using commonly used signal quality descriptors, including the first analysis of linearity and crosstalk. Our investigation contributes to a growing body of literature that characterizes the nature of eye tracking data in AR environments. Based on our findings, we recommend future studies using the HoloLens 2 include a recalibration task that captures gaze data across the viewing field to enable post hoc data correction. 

\section{Acknowledgment}
This material is based upon work supported by the National Science Foundation Graduate Research Fellowship under Grant No. DGE-1840989.

\bibliographystyle{unsrt}
\bibliography{main}
\end{document}